\documentclass[pra,twocolumn,showpacs,amsmath,amssymb]{revtex4}
\usepackage{graphicx}% Include figure files
\usepackage{dcolumn}% Align table columns on decimal point
\usepackage{bm}% bold math

\begin{document}

\title{Thermodynamic properties of two-component fermionic atoms trapped in a two-dimensional optical lattice 
%:an application of the self-energy functional approach to inhomogeneous systems
}

\author{Kensuke Inaba}
\author{Makoto Yamashita}%

%\inst{%
%Department of Applied Physics, Osaka University, Suita, Osaka 565-0871 %, Japan
%}%
\affiliation{
NTT Basic Research Laboratories, NTT Corporation, Atsugi 243-0198, Japan
}
\affiliation{
CREST, JST, Chiyoda-ku, Tokyo 102-0075, Japan
}

\date{\today}% It is always \today, today,

%\abst{
\begin{abstract}
We study the finite temperature properties of two-component fermionic atoms trapped in a two-dimensional optical lattice.
We apply the self-energy functional approach to the two-dimensional Hubbard model with a harmonic trapping potential, and systematically investigate the thermodynamic properties of this system.
We find that entropy and grand potential provide evidence of a crossover between the Mott insulating and metallic phases at certain temperatures.
In addition, we find that entropy exhibits a cusp-like anomaly at lower temperatures, suggesting a second or higher order antiferromagnetic transition.
We estimate the antiferromagnetic transition temperatures, and clarify how the trapping potential affects this magnetic transition.
\end{abstract}
%}


%\kword{orbital-selective Mott transition, self-energy functional approach}
%Use showkeys class option if keyword
\preprint{APS/123-QED}

\maketitle

%%%%-------------------------------------------------------------%%%%
%%%%%　　　%%　　　%%%　　　%%　　　　%%　%%%　　　　%%%　%%%%　%%%%%
%%%%%　%%%%%%　%%%%%%　%%%%%%%%%%　%%%%%　%%%　%%%%　%%%　　%%　%%%%%
%%%%%%　　%%%　　　%%　%%%%%%%%%%　%%%%%　%%%　%%%%　%%%　%　%　%%%%%
%%%%%%%%%　%%　%%%%%%　%%%%%%%%%%　%%%%%　%%%　%%%%　%%%　%%　　%%%%%
%%%%%　　　%%　　　%%%　　　%%%%%　%%%%%　%%%　　　　%%%　%%%%　%%%%%
%%%%-------------------------------------------------------------%%%%
\section{introduction}
%%%===%%%  i1  分野の特定
Ultracold atoms in an optical lattice created by interfering laser beams are bridging the gap between theoretical and experimental studies in the field of condensed matter physics.
We can control the depth, interstice and dimension of the lattice by manipulating lasers, and the interaction between ultracold atoms by using the Feshbach resonance \cite{Chen2005}.
This high degree of controllability allows us to simulate correlated many-body systems experimentally \cite{Regal2004,Bloch2005,Greiner2008,Jo2009}.
%Ultracold atoms in an optical lattice created by interfering laser beams are providing significant progress in the field of condensed matter physics \cite{Greiner2008,Bloch2005}.
Theoretically, it has been pointed out that this system can be regarded as a realization of the Hubbard model, which is one of the most fundamental models including the many-body effects \cite{Jaksch1998,Zwerger2003,Jaksch2005}.
In fact, various phenomena described by this model have been successfully demonstrated in experiments \cite{Greiner2002,Kohl2005,Jordens2008,Schneider2008,Gemelke2009,Fukuhara2007,Fukuhara2009}.
%It is thus expected that the optical lattice systems bridge the distance between the theory and experiment.
%%%===%%%  i2  過去の研究

The first experimental progress was reported as an observation of the quantum phase transition of bosonic $^{87}$Rb atoms between a superfluid and a Mott insulator in an optical lattice \cite{Greiner2002}.
%Bring our attention to
As regards the fermionic atoms, the Fermi surface and its topological change were observed in the fermionic $^{40}$K optical lattice system, suggesting a metal to band-insulator transition \cite{Kohl2005}.
%A disappearance of coherent interference was observed for fermionic $^6$Li atoms \cite{Chin2006}, indicating a transition from a superfluid state to an insulating state.
%However, there are still some controversies as to what type of insulator appears in this experiment \cite{Zhai2007,Moon2007,Chien2008,Higashiyama2008,Burkov2009}.
Recently, by combining numerical and experimental studies, it was clarified that a Mott transition was realized in $^{40}$K optical lattice systems \cite{Jordens2008,Schneider2008}.
In these reports, the Hubbard model with a trapping potential was analyzed by using a local density approximation with the dynamical mean-field approach (LDA+DMFT)\cite{Schneider2008} and a zero-tunneling limit calculation \cite{Jordens2008}.
%These numerical calculations have been done for the Hubbard model with a trapping potential by means of the local density approximation with the dynamical mean-field approach (LDA+DMFT)\cite{Schneider2008} and the zero-tunneling limit calculation \cite{Jordens2008}.
The effects of the trapping potential, which is an important characteristic of ultracold atoms in an optical lattice, have been properly taken into account in addition to those of the correlations.
However, these calculations present certain difficulties in terms of investigating magnetic transitions.
%not properly dealt with the tunneling effects. As a result, there are using these methods
%%%===%%%  i3  問題点（過去との比較）
The observation of the magnetic ordered phase in the optical lattice system is a major concern for condensed matter physicists, because it could provide ways to elucidate the nature of high $T_c$ superconductors \cite{Lee2006}.
%We thus need theoretical studies about the magnetic transitions of the Hubbard model in an optical lattice.
%The Hubbard model with a trapping potential has been investigated at zero temperature by the recently developed real-space dynamical mean-field theory (R-DMFT) \cite{Snoek2008}.
By using the recently developed real-space dynamical mean-field theory (R-DMFT) \cite{Helmes2008,Snoek2008,Koga2008,Koga2009}, it has been pointed out that an antiferromagnetic (AF) ordered phase is stable in the Hubbard model with a trapping potential at zero temperature \cite{Snoek2008}.
This naturally motivates us to undertake a detailed study of both the magnetic transition and the Mott transition at finite temperatures using a reliable numerical method.
%However, there are far less systematic studies at finite temperatures.
%It is desirable to investigate the finite temperature properties of the magnetic transitions in this system.
%Furthermore, for comprehensive understandings, we thus need systematic numerical studies in which both the magnetic transition and Mott transition are precisely investigated at finite temperatures.
%%%===%%%  i4  研究内容
%For comprehensive understandings of the properties of optical lattice systems, 

%We thus need  theoretical studies which appropriately treat the effects of a trapping potential and strong correlations, and thermal fluctuations.
For this purpose, we investigate the two-component Fermi-Hubbard model in a two-dimensional (2D) optical lattice with a harmonic trapping potential at finite temperatures.
We make use of the self-energy functional approach (SFA), which has been successfully applied to homogeneous Hubbard-type lattice models \cite{Potthoff2003,Potthoff2003a,Potthoff2003c,Balzer2009}.
We demonstrate that this method can properly take account of the effects of the trapping potential in addition to those of strong correlations and thermal fluctuations.
SFA further provides us with important thermodynamic quantities such as grand potential and  entropy.
%It is difficult to systematically investigate these quantities by other methods, such as LDA+DMFT\cite{Schneider2008} and R-DMFT\cite{Snoek2008}. %, which correctly deal with the effects of correlations and confinement.
%Since thermodynamic quantities such as the grand potential and the entropy have much information about the finite temperature properties of atoms, we systematically investigate the behavior of these quantities of the present model.
We find that both Mott and AF transitions can be characterized by the behavior of these thermodynamic quantities.
The AF transition temperatures of the present model are systematically examined by varying the temperature, interaction strengths and curvatures of the harmonic trapping potential.
%In addition, by systematic calculation with varying temperature, the interaction strength, and the curvature of the trapping potential, we estimate the AF transition temperatures of the present model.
%Finally, we propose the optimized parameter region to observe the magnetic ordered state.

%%%===%%%  i5  論文構成
This paper is organized as follows. 
In Sec. \ref{sec_model}, we introduce the two-component Fermi-Hubbard model on a 2D optical lattice with a harmonic trapping potential. 
In Sec. \ref{sec_method}, we briefly outline the application of SFA to the present system.
In Sec. \ref{sec_Mott}, we discuss the finite temperature properties of the Mott insulating region.
In Sec. \ref{sec_AF}, we discuss how the AF ordered region is realized in the system, and estimate the AF transition temperatures.
In Sec. \ref{sec_summary}, we briefly summarize this paper.

%%%%-------------------------------------------------------------%%%%
%%%%%　　　%%　　　%%%　　　%%　　　　%%　%%%　　　　%%%　%%%%　%%%%%
%%%%%　%%%%%%　%%%%%%　%%%%%%%%%%　%%%%%　%%%　%%%%　%%%　　%%　%%%%%
%%%%%%　　%%%　　　%%　%%%%%%%%%%　%%%%%　%%%　%%%%　%%%　%　%　%%%%%
%%%%%%%%%　%%　%%%%%%　%%%%%%%%%%　%%%%%　%%%　%%%%　%%%　%%　　%%%%%
%%%%%　　　%%　　　%%%　　　%%%%%　%%%%%　%%%　　　　%%%　%%%%　%%%%%
%%%%-------------------------------------------------------------%%%%
\section{model}\label{sec_model}
\begin{figure}[t]
\includegraphics[width=2.5cm]{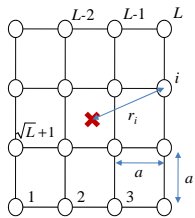}
\caption{(Color online) A schematic diagram of the 2D lattice, where the cross is the center of the harmonic trapping potential, $L$ is the total number of sites, $a$ is the lattice distance, and $r_i$ is the distance between the center of the trap and the $i$th site.
The numbers written in the top right of the lattice sites correspond to the site indices.
%Some of the indices of a site are shown at upper right of the corresponding site.
}
\label{fig_lattice}
\end{figure}

%%%===%%%  m1  モデルの導入
%The properties of fermionic atoms in an optical lattice can be well described by the Hubbard model with a trapping potential \cite{Jaksch1998,Zwerger2003,Jaksch2005}.
In this paper, we investigate the two-component Hubbard model on a 2D lattice with a trapping potential that we assume to be harmonic.
The Hamiltonian of this model is given by ${\cal H}={\cal H}_{\bf t}+{\cal H}_{\bf U}$, 
%%%%%%%%%%%%%%%%%%%%%%%%%%%%%%%
\begin{eqnarray}
{\cal H}_{\bf t}&=&J \sum_{<i,j>}\sum_{\sigma}c^\dag_{i\sigma} c_{j\sigma}+
\sum_i\sum_\sigma (V_t r_i^2-\mu)n_{i\sigma}, \\
%t_{ij}
{\cal H}_{\bf U}&=&
\sum_{i} U n_{i \uparrow} n_{i \downarrow}
\label{eq_model},
\end{eqnarray}
%%%%%%%%%%%%%%%%%%%%%%%%%%%%%%%%%%%%%%%%%%
where $c^\dag_{i\sigma} (c_{i\sigma})$ creates (annihilates) a fermionic atom with pseudospin $\sigma(=\uparrow, \downarrow)$ at the $i$th site, and $n_{i\sigma}=c^\dag_{i\sigma}c_{i\sigma}$.
Here, we describe the nearest-neighbor hopping integral as $J$, the curvature of a harmonic trapping potential as $V_t$, the chemical potential as $\mu$, and the interaction strength between two atoms with different pseudospins as $U$. 
%Here, $L$ is the total number of sites of the 2D lattice.
For simplicity, we define the site index $i$ as shown in Fig. \ref{fig_lattice}, where $L$ is the number of sites, $a$ is the lattice distance, and $r_i$ is the distance between the $i$th site and the center of the trap.
%For convenience of following explanation about the method used in this paper, 
It is useful to introduce the parameter matrix $\bf t$, which characterizes the non-interacting Hamiltonian $\cal H_{\bf t}$ as
\begin{eqnarray}
\left[{\bf t}\right]_{ij}&=
\left\{\begin{array}{lc}
V_t r_i^2 & \text{ if $i$ = $j$ } \\
 J & \text{if the $i$th site neighbors the $j$th site} \\
0 & \text{otherwise}
\end{array}\right..
\end{eqnarray}

%...................................
%...................................

%%%===%%%  m2 モデルの詳細： スケーリング
We define the scales of length and energy in the following way such that the calculated results do not depend on the details of the parameters, {\it e.g.} $L, a, V_t$, and $\mu$.
We introduce the characteristic trap length $r_t=\sqrt{N_\sigma}/\sqrt{\pi}a$, which corresponds to the radius of a non-interacting atomic cloud in the zero-tunneling limit, where $N_\sigma$ is the total number of atoms with pseudospin $\sigma$.
% ($N_\uparrow = N_\downarrow$).
The distance $r_i$ is rescaled by this $r_t$ as $r_{SC}\equiv r_i/r_t$.
We also introduce the bandwidth $W=8J$ as a scale of energy.
We consider the characteristic trap energy $E_t =V_t r_t^2$, which corresponds to the Fermi energy in the zero-tunneling limit \cite{Schneider2008}.
In addition to $E_t/W$, we consider two parameters:
the temperature $T/W$ and the interaction strength $U/W$. 
%In the following, we renormalize the energy scales $E_t, T$ and $U$ with $W$ , and rescale the distance $r_i$ with $r_t$.

%%%===%%%  m3  物理量
Before concluding this section, we summarize the physical quantities calculated in this paper.
% in order to systematically investigate the quantum phases of the present system.
The rescaled cloud size, which characterizes the Mott
 and band insulating phases \cite{Schneider2008}, is defined as,
\begin{align}
R_{SC}&=\langle r\rangle/r_t, \\
\langle r\rangle &= \sqrt{\sum_{i} r_i^2N_{i}} / \sqrt{N_{\rm tot}}, 
\end{align}
where $N_i=\sum_\sigma \langle n_{i\sigma}\rangle$ is the number of atoms at the $i$th site, and $N_{\rm tot}=\sum_{i} N_i$ is the total number of atoms.
To discuss the magnetic ordered phase, we calculate the magnetization $M_i=\langle n_{i\uparrow}\rangle-\langle n_{i\downarrow}\rangle$.
We also investigate thermodynamic quantities: grand potential $\Omega$ and entropy $S=-\partial \Omega /\partial T$, which are sensitive to both the Mott transition and the AF transition.

\section{method}\label{sec_method}
%%%===%%%  t1  方法の導入 SFA
The thermodynamic properties of the present system are studied with the self-energy functional approach (SFA) \cite{Potthoff2003,Potthoff2003a,Potthoff2003c}, which is based on the Luttinger-Ward variational method \cite{Luttinger1960}.
It has been pointed out that SFA allows us to undertake an efficient investigation of the finite temperature properties of homogeneous Hubbard-type lattice systems, 
for instance, the infinite-dimensional Hubbard model \cite{Potthoff2003,Potthoff2003a} and also the 2D Hubbard model \cite{Balzer2009}, taking account of the effects of strong correlations.
We extend this method to deal with inhomogeneous systems.
Here, we explain an application of this SFA to the Hubbard model with a trapping potential.

%%%===%%%  t2  方法の説明 SFA
% of the Hubbard type lattice system with the Hamiltonian ${\cal H}={\cal H}_{\bf t}+{\cal H}_{\bf U}$
First, we begin by briefly explaining the general framework of the SFA.
According to the Luttinger-Ward functional approach \cite{Luttinger1960}, the grand potential $\Omega$ of a Hubbard-type system described by the Hamiltonian ${\cal H}={\cal H}_{\bf t}+{\cal H}_{\bf U}$ is written as,
%------------------------   grand potential   -----------------------
\begin{eqnarray}
\Omega [{\boldsymbol \Sigma}]&=&F[{\boldsymbol \Sigma}] + 
{\rm Tr}\ln [({\bf G}_0^{-1}-{\boldsymbol \Sigma})^{-1}]\label{eq_LW},
\end{eqnarray}
%--------------------------------------------------------------------
where $F[{\boldsymbol \Sigma}]$ is the Legendre transformation of the Luttinger-Ward potential, ${\boldsymbol \Sigma}$ is the self-energy, and ${\bf G}_0=(\omega - \mu - {\bf t})^{-1}$ is the non-interacting Green function.
Here, we use the notation ${\rm Tr}{\bf A}=T\sum_{\omega_n, i}\left[{\bf A}\right]_{ii}(i\omega_n)$, where $\omega_n=(2n+1)\pi T$ is the Matsubara frequency.
Under the condition ${ \partial \Omega[{\boldsymbol \Sigma}] }/{ \partial {\boldsymbol \Sigma}} =0$, we obtain the physical Green function $\bf G$ that satisfies the Dyson equation ${\bf G}^{-1}={\bf G}_0^{-1} -{\boldsymbol \Sigma}$. 
%\reviseAdd[(1)]{
In general, the functional $F[{\boldsymbol \Sigma}]$ is not known explicitly, 
which prevents an evaluation of $\Omega[{\boldsymbol \Sigma}]$. 
However, the potential
%}
$F[{\boldsymbol \Sigma}]$ does not depend on the details of the non-interacting Hamiltonian ${\cal H}_{\bf t}$ as long as the shape of the interaction term ${\cal H}_{\bf U}$ remains unchanged \cite{Potthoff2003}.
This allows us to introduce a reference system that has a Hamiltonian with the same interaction term as that of the original system. 
This reference Hamiltonian is explicitly given by ${\cal H}^{\rm ref}={\cal H}_{\bf t'}+{\cal H}_{\bf U}$ with the {\it variational} parameter matrix ${\bf t}'$.
The grand potential of the reference system $\Omega^{\rm ref}$ is then written as,
\begin{eqnarray}
\Omega^{\rm ref} [{\boldsymbol \Sigma}]&=&F[{\boldsymbol \Sigma}] + 
{\rm Tr}\ln [({\bf G}_0^{'-1}-{\boldsymbol \Sigma})^{-1}]\label{eq_LWref},
\end{eqnarray}
where ${\bf G}'_0=(\omega - \mu - {\bf t}')^{-1}$.
By subtracting Eq. (\ref{eq_LWref}) from Eq. (\ref{eq_LW}),
we can rewrite the grand potential of the original system as a function of the self-energy for the reference system  ${\boldsymbol \Sigma}^{\rm ref}$:
%--------------------------------------------------------------------
\begin{eqnarray}
\Omega [{\boldsymbol \Sigma}^{\rm ref}]&=&\Omega^{\rm ref}
%%%({\bf t^\prime})
    +{\rm Tr}\ln
    \left[(\omega+\mu-{\bf t}-{\boldsymbol \Sigma}^{\rm ref})^{-1}\right]
    \nonumber\\
    &-&{\rm Tr}\ln
    \left[(\omega+\mu-{\bf t}'-{\boldsymbol \Sigma}^{\rm ref})^{-1}\right].\label{eq:omega_SFA}
\end{eqnarray}
%--------------------------------------------------------------------
A reference system with an optimized parameter matrix ${\bf t}'$ satisfying the condition 
\begin{equation}
\partial \Omega[{\boldsymbol \Sigma^{\rm ref}}]/\partial {\bf t}'=0,
\label{eq_v_condition}
\end{equation}
gives us an appropriate self-energy ${\boldsymbol \Sigma}^{\rm ref}$, Green function ${\bf G}=(\omega-\mu-{\bf t}-{\boldsymbol \Sigma}^{\rm ref})^{-1}$ and grand potential $\Omega[\boldsymbol{\Sigma}^{\rm ref}]$, which approximately describe the physical quantities of the original system. 
% with Hamiltonian ${\cal H}={\cal H}_{\bf t}+{\cal H}_{\bf U}$.

An application of the SFA to the Hubbard model with a trapping potential is achieved as follows.
The simplest reference system with which to investigate the present model is $L$-sets of two-site Anderson impurity models \cite{Potthoff2003a}. 
Here, the $i$th impurity site, corresponding to the $i$th site of the original lattice, is connected to the $i$th non-interacting atomic bath. 
The Hamiltonian of this reference system is given by the form: ${\cal H}^{\rm ref}=\sum_i{\cal H}^{\rm ref}_{i}$, and
%%%%%%%%%%%%%%%%%%%%%%%%%%%%%
\begin{eqnarray}
%---------------    Hamiltonian of reference system    --------------
&&{\cal H}^{\rm ref}_{i}=\sum_{\sigma} 
(\epsilon_{i}+\sigma h_i)c^\dag_{i\sigma} c_{i\sigma}\nonumber\\
%\epsilon^a_{i\sigma}a^{\dag}_{i\sigma}a_{i\sigma}\right]\nonumber\\
&&        +\sum_{\sigma}\left(V_{i\sigma}
          c^\dag_{i\sigma}a_{i\sigma}+H.c.\right)
+U n_{i \uparrow} n_{i \downarrow},
\label{eq_ref_model}
\end{eqnarray}
%%%%%%%%%%%%%%%%%%%%%%%%%%%%%%%%%%
where $a^{\dag}_{i\sigma} (a_{i\sigma})$ creates (annihilates) an atom with pseudospin $\sigma$ at the $i$th atomic bath.
The variational parameter $V_{i\sigma}$ is the hybridization of the impurity and the atomic bath, $\epsilon_{i}$ is the effective potential, and $h_i$ is the effective magnetic field.
Here, we briefly explain the role of variational parameters $\epsilon_{i}$, $h_i$ and $V_{i\sigma}$.
By optimizing an effective potential $\epsilon_{i}$ under the condition $\partial\Omega/\partial \epsilon_i=0$, the number of atoms $N_i$ is properly adjusted \cite{Potthoff2003c}.
%the physical relation $\langle n_{i\sigma}\rangle =T\sum_{\omega_n}\left[{\bf G}_\sigma(i\omega_n)\right]_{ii}$ can be reproduced.
We can discuss the stability of magnetic ordered phases via the condition $\partial\Omega/\partial h_i=0$ \cite{Potthoff2003c,Dahnken2004,Balzer2008}.
The hybridization $V_{i\sigma}$ effectively describes the hopping integral between the $i$th site and adjacent sites in the original lattice; 
therefore, we can also discuss the Mott transition via the conditions $\partial \Omega/\partial V_{i\sigma}=0$ \cite{Potthoff2003,Potthoff2003a}.
Details of the role of the atomic bath with the hybridization term will be discussed in appendix A.

By employing exact diagonalization, we can easily obtain the grand potential $\Omega^{\rm ref}_i$ and the self-energy $\Sigma^{\rm ref}_{i\sigma}$ of the $i$th reference system.
Now, the grand potential of the original system is given by,
%--------------------------------------------------------------------
\begin{align}
\Omega=\sum_i&\Omega^{\rm ref}_i %-2LT\ln2 
-\sum_{\sigma}{\rm Tr}\ln {\bf G}^{\rm ref}_{\sigma}
+\sum_{\sigma}{\rm Tr}\ln{\bf G}_{\sigma},
\end{align}
\begin{align}
{\bf G}_{\sigma}&=(i\omega_n+\mu-{\bf t}-{\boldsymbol \Sigma}_\sigma^{\rm ref})^{-1}, \\
{\bf G}^{\rm ref}_{\sigma}&=( 
      {\bf G}_{0\sigma}^{\rm ref -1}
      -{\boldsymbol \Sigma}^{\rm ref}_{\sigma})^{-1},
\label{eq_omega_SFA}
\end{align}
where $[{\bf G}_{0\sigma}^{\rm ref}]_{ij}=\delta_{ij}/(i\omega_n-\epsilon_{i}-\sigma h_i-V_{i\sigma}^2/i\omega_n)$, and $\left[{\boldsymbol \Sigma}_\sigma^{\rm ref}\right]_{ij}=\delta_{ij}\Sigma^{\rm ref}_{i\sigma}$.
The variational condition (\ref{eq_v_condition}) is rewritten as,
\begin{eqnarray}
%\frac{\partial \Omega}{\partial {\bf t}'_i}= 
T\sum_{i\omega_n}
\left(\left[{\bf G}_{\sigma}\right]_{ii}-
\left[{\bf G}^{\rm ref}_{\sigma}\right]_{ii}\right)
\frac{\partial\Sigma^{\rm ref}_{i\sigma}}{\partial {\bf t}'_i} = 0, 
\label{eq_euler}
\end{eqnarray}
where we denote the variational parameter sets as ${\bf t}'_i=\{\epsilon_i, h_i, V_{i\uparrow}, V_{i\downarrow}\}$.
Note that $\left[{\bf G}^{\rm ref}_{\sigma}\right]_{ii}$ and $\partial\Sigma^{\rm ref}_{i\sigma}/\partial {\bf t}'_i$ depend only on one parameter set ${\bf t'}_i$.
In contrast, $\left[{\bf G}_{\sigma}\right]_{ii}$ depends on all of the elements of the variational parameter matrix $\bf t'$, and we have to solve an equation of the $4L$th degree.
It is still difficult for the large size system to solve equation (\ref{eq_euler}) and to optimize the variational parameter matrix ${\bf t}'$.
To avoid this difficulty, we introduce another parameter matrix ${\bf t}^*$ and local Green functions as $G^{\rm loc}_{i\sigma}({\bf t}', {\bf t}^*)=\left({G}^{\rm cav -1}_{i\sigma}({\bf t}^*)-\Sigma_{i\sigma}^{\rm ref}({\bf t}')\right)^{-1}$, where $G^{\rm cav}_{i\sigma}({\bf t}^*)=\left(\Sigma^{\rm ref}_{i\sigma}({\bf t}^*)+\left[{\bf G}_{\sigma}({\bf t}^*)\right]_{ii}^{-1}\right)^{-1}$.
%, here we explicitly describe the dependence of Green functions on the parameter matrices.
%We replace $\left[{\bf G}_{\sigma}\right]_{ii}({\bf t}')$ with $G^{\rm loc}_{i\sigma}({\bf t}', {\bf t}^*)$ considering the additional self-consistent conditions $\left[{\bf G}_{\sigma}\right]_{ii}({\bf t}')=G^{\rm loc}_{i\sigma}({\bf t}', {\bf t}^*)$.
We replace $\left[{\bf G}_{\sigma}\right]_{ii}$ in equation (\ref{eq_euler}) with $G^{\rm loc}_{i\sigma}$ considering the additional self-consistent condition $\left[{\bf G}_{\sigma}\right]_{ii}=G^{\rm loc}_{i\sigma}$.
If ${\bf t}^*$ is given, the variational condition (\ref{eq_euler}) is finally rewritten as,
\begin{align}
%\frac{\partial \Omega}{\partial {\bf t}'_i}
%&=T\sum_{i\omega_n}
%\left(G_{i\sigma}^{\rm loc-1}-
%{G}^{\rm ref}_{i\sigma}\right)
%\frac{\partial\Sigma^{\rm ref}_{i\sigma}}{\partial {\bf t}'_i} \nonumber\\
&\frac{\partial}{\partial {\bf t}'_i}
\left(\Omega^{\rm ref}_i-T\sum_{\sigma\omega_n}\ln G^{\rm ref}_{i\sigma}
+T\sum_{\sigma\omega_n}\ln G^{\rm loc}_{i\sigma} \right)=0.
\label{eq_euler_}
\end{align}
We note that the equation of the $4L$th degree is decomposed into independent $L$-sets of the $4$th degree.
Now, we choose ${\bf t}^*$ as a parameter matrix satisfying this new condition (\ref{eq_euler_}), and repeatedly solve this variational problem until ${\bf t}^*$ converges (at the same time, the self-consistent condition $\left[{\bf G}_{\sigma}\right]_{ii}=G^{\rm loc}_{i\sigma}$ is satisfied).
In summary, we self-consistently solve independent $L$-sets of easily solvable SFA problems instead of a huge problem with the $4L$th degree.

The procedure mentioned above can be extended to any other inhomogeneous systems.
In addition, if we deal with the attractively interacting systems, we can straightforwardly extend the scope of this method to the $s$-wave superfluid phase by adding the term, $\Delta_i c^\dag_{i\uparrow}c^\dag_{i\downarrow}+{\rm H.c.}$, to the reference Hamiltonian (\ref{eq_ref_model}), where $\Delta_i$ is a variational parameter corresponding to the superfluid order parameter.
Furthermore, the SFA has been successfully applied to multi-component Fermi systems \cite{Inaba2009a}, where it has been suggested that novel quantum phase transitions could be observed \cite{Rapp2007,Wilczek2007,Cherng2007,Cazalilla2009}. 
Promising candidates for such systems are fermionic $^{173}$Yb atoms \cite{Fukuhara2007} and/or $^6$Li atoms \cite{Ottenstein2008,Huckans2009}.
%Such systems may be realized by fermionic $^{173}$Yb atoms \cite{Fukuhara2007} and/or $^6$Li atoms \cite{Ottenstein2008,Huckans2009} in the near future.

%%%===%%%  t3  物理量など SFA
Now, we describe briefly how to calculate the physical quantities.
It is possible to evaluate various quantities denoted in the previous section by means of the Green function ${\bf G}(\omega)$, the grand potential $\Omega$, and its derivatives.
For instance, the total number of atoms is obtained by the relations: $N_{\rm tot}=\partial \Omega/\partial \mu$ or $N_{\rm tot}= {\rm Tr} {\bf G}$.
Note that the variational condition $\partial \Omega/\partial {\bf t'}=0$ guarantees that these two relations are equivalent to each other \cite{Ortloff2007}.
In addition, the SFA allows us to calculate quantities in momentum space, {\it e.g.} experimentally observed time of flight images.
We can also calculate angle resolved photo emission spectra using the present scheme, which is closely related to a recent observation by the JILA group in Ref. \cite{Stewart2008}.
However, it is beyond our current scope to investigate these quantities in momentum space. 
%left as future works.
%$A(\omega)=-{\rm Im}{G}_{\bf k}(\omega+i0)/\pi$, where $G_{\bf k\sigma}$ is the Green function in the wavenumber space.
%time of flight images $n_{\bf k}=\sum_nG_{{\bf k}\sigma}(i\omega_n)$ and

%%%===%%%  t4  関連した方法 SFA
We close this section with few comments on the superiority of the SFA to other related numerical approaches.
The zero-tunneling limit calculation and LDA+DMFT present certain difficulties when we discuss quantum phase transitions accompanied by spontaneous symmetry breaking.
Although R-DMFT can be applied to the magnetic ordered phase \cite{Snoek2008} and the superfluid phase \cite{Koga2008,Koga2009}, it is difficult to investigate the finite temperature properties systematically and precisely with this method.
%%%===%%%  t5  コメント SFA
In contrast, the SFA provides comprehensive ways to investigate various quantum phase transitions at finite temperatures.
Moreover, this method allows us to calculate various useful quantities observed experimentally.
%Such systematic studies are difficult for other methods such as the zero-tunneling limit calculation, LDA+DMFT and R-DMFT.

%%%%-------------------------------------------------------------%%%%
%%%%%　　　%%　　　%%%　　　%%　　　　%%　%%%　　　　%%%　%%%%　%%%%%
%%%%%　%%%%%%　%%%%%%　%%%%%%%%%%　%%%%%　%%%　%%%%　%%%　　%%　%%%%%
%%%%%%　　%%%　　　%%　%%%%%%%%%%　%%%%%　%%%　%%%%　%%%　%　%　%%%%%
%%%%%%%%%　%%　%%%%%%　%%%%%%%%%%　%%%%%　%%%　%%%%　%%%　%%　　%%%%%
%%%%%　　　%%　　　%%%　　　%%%%%　%%%%%　%%%　　　　%%%　%%%%　%%%%%
%%%%-------------------------------------------------------------%%%%
\section{Mott insulating phase}\label{sec_Mott}
%%%===%%%  r1  結果の導入 モット絶縁体について
%...................................
\begin{figure}[t]
\includegraphics[width=8cm]{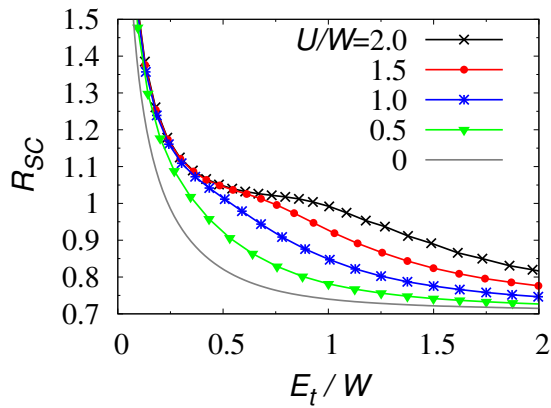}
\caption{(Color online) Rescaled cloud size $R_{SC}$ as a function of $E_t/W$ for different $U/W$ at $T/W=0.1$.}\label{fig_rsc}
\end{figure}
%...................................
%...................................
\begin{figure}[t]
\includegraphics[width=8cm]{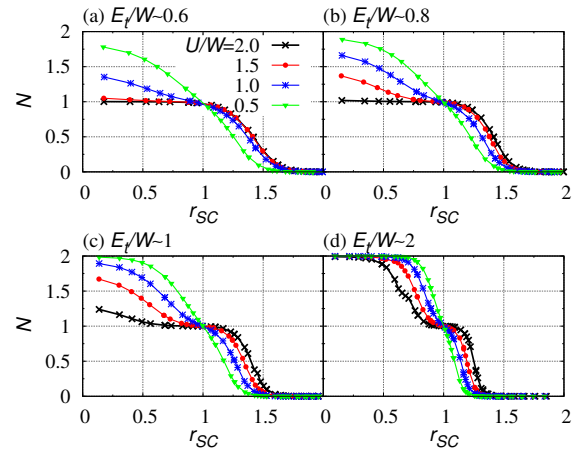}
\caption{(Color online) Density profiles of atomic cloud, {\it i.e.} the number of atoms $N_i$ as a function of the rescaled distance $r_{SC}$ for several fixed values of $E_t/W$: (a) $0.6$, (b) $0.8$, (c) $1.0$ and (d) $2.0$.
The interaction strength is varied from $U/W=0.5$ to $U/W=2.0$. 
The temperature is fixed at $T/W=0.1$.}\label{fig_num}
\end{figure}
%...................................
In this section, we investigate the finite temperature properties of the 2D Hubbard model with a harmonic trapping potential.
In particular, we focus on the behavior of the thermodynamic quantities of this model at certain temperatures, where the Mott insulating region is observed.
%First, we calculate the quantities previously investigated in Ref. \cite{Schneider2008}, and verify SFA is successfully applied to the system with a trapping potential.
%Next, we investigate thermodynamic quantities such as the grand potential and the entropy.
%We demonstrate that the behavior of thermodynamic quantities show a characteristic of the crossover between metallic to Mott insulating phases.
%Finally, we discuss the effect of thermal fluctuations to the properties of the Mott insulating region.

First, to discuss the validity of our method, we calculate the rescaled cloud size $R_{SC}$ which was previously evaluated for the three dimensional (3D) model by the LDA+DMFT approach in Ref. \cite{Schneider2008}.
Figure \ref{fig_rsc} shows $R_{SC}$ as a function of $E_t/W$ for different $U/W$ values, where the temperature is fixed at $T/W=0.1$. 
%%%===%%%  r2  発見の提示   モット絶縁体について　Rsc(Et) N(r) with U
When the interaction strength is small ($U/W=0, 0.5$ and $1.0$), $R_{SC}$ decreases rapidly in the weakly trapped region $E_t/W<1$, while it changes little in the strongly trapped region $E_t/W>1$.
It should be noted that, in the strongly trapped limit $E_t/W \gg 1$, $R_{SC}$ finally converges to $1/\sqrt{2}\sim0.7$, which corresponds to the cloud size of non-interacting atoms without tunneling.
This indicates that most of the atoms except for those around the site $r_i\sim r_t$ become inactive; in other words, the band-insulating like state appears \cite{Schneider2008}.
On the other hand, for large $U/W$ values of $1.5$ and $2$, the $R_{SC}$ curves exhibit shoulder like structures in the region $0.4\lesssim E_t/W \lesssim 0.7$ and $0.4\lesssim E_t/W \lesssim 1.0$, respectively, suggesting the formation of the Mott insulating state there.

%%%===%%%  r1  結果の導入 モット絶縁体：プロファイル
To clarify these points further, we calculate the density profiles, namely the number of atoms $N_i$ as a function of rescaled distance $r_{SC}=r_i/r_t$, for several fixed values of $E_t/W$ and $U/W$.
We show the results in Fig. \ref{fig_num}.
Since we adopt a grand canonical ensemble in our calculations, the $E_t/W$ values in Fig. \ref{fig_num} have a small deviation of about $\pm0.05$.  %(\propto N_{\rm tot})
%%%===%%%  r2  結果の導入 モット絶縁体：プロファイル
In Fig. \ref{fig_num}(a), at $E_t/W\sim0.6$, a well developed Mott plateau with $N_i=1$ from the center to the edge of the cloud appears for large $U/W$ values of $1.5$ and $2.0$.
These plateau profiles are gradually deformed as $E_t/W$ is increased. 
From Fig. \ref{fig_num}(b) and (c), we see that the additional atoms are loaded in the Mott plateau around the center at $E_t/W\sim0.8$ ($E_t/W\sim1.0$) when the interaction strength is $U/W=1.5$ ($2.0$). 
In Fig. \ref{fig_num}(c) and (d), the Mott plateau is not dominant and is limited to the shell region around $r_{SC}\sim 1$ even for the large interaction $U/W=2.0$.
%Consequently, we can see the shell structure of the Mott plateau around $r_{SC}\sim 1$ for the strongly trapped region ($E_t \gtrsim 1$).
In all the panels in Fig. \ref{fig_num}, for small $U/W$ values of $0.5$ and $1.0$, there are no Mott plateaus over the entire $E_t/W$ range.
This density profile behavior is consistent with the shoulder structure of the $R_{SC}$ curves in Fig. \ref{fig_rsc}.

%For any $U/W$ and $E_t/W$, the density of atom keeps one at the Fermi length of the zero-tunneling and non-interacting limit, $N_i(r_{SC}=1)\sim1$. 
%It is due to the fact that the volume of the Fermi surface is always conserved.

%%%===%%%  r3  関連した研究 モット絶縁体について
Our results for the 2D system exhibit good qualitative agreement with previous theoretical and experimental results for the 3D system reported in  Ref. \cite{Schneider2008}. 
This suggests that a proper renormalization allows us to compare the results between two- and three-dimensional optical lattice systems qualitatively.
%定量的にも多少はいける。

%%%===%%%  r1  結果の導入 モット絶縁体について　熱力学量
%...................................
\begin{figure}[t]
\includegraphics[width=8cm]{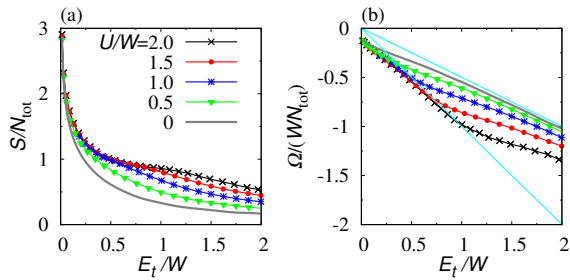}
\caption{(Color online) (a) Entropy per atom $S/N_{\rm tot}$ and (b) grand potential per atom $\Omega/N_{\rm tot}$ for several values of $U/W$ at $T/W=0.1$. Thin (light blue) lines  in the panel (b) mean $\Omega=-E_t N_{\rm tot}/2$ and $\Omega=-E_t N_{\rm tot}$ from top to bottom.}\label{fig_ent}
\end{figure}
%...................................
As a complementary study, we next investigate other thermodynamic quantities at a fixed temperature $T/W=0.1$.
We calculate the entropy per atom $S/N_{\rm tot}$ and the grand potential per atom $\Omega/N_{\rm tot}$ as shown in Fig. \ref{fig_ent}(a) and (b), respectively.
%%%===%%%  r2  発見の提示   モット絶縁体について　熱力学量　S(r) with Et
Curves of $S/N_{\rm tot}$ reveal a similar $E_t/W$ dependence of $R_{SC}$ to that in Fig. \ref{fig_rsc}.
For strongly interacting cases such as $U/W=1.5$ and $2.0$, the shoulder structure appears in the $S/N_{\rm tot}$ curves around $E_t/W\sim 0.5$, suggesting that the strong correlations induce localized free spins whose entropy takes a constant value of $\ln2$.
%In the Mott insulating region, doubly occupied and unoccupied sites are far less than the singly occupied sites.
%For the Mott insulating region, the shoulder structure thus appears in the curve of $S/N_{\rm tot}$.
It is reasonable to expect that a magnetic ordered phase will appear because these free spins are interacting with each other.
However, at $T/W=0.1$, thermal fluctuations will destroy such an ordered phase, as we discuss in the next section.
% (the heat capacity $C=T\partial S/\partial T$ of such atoms is zero).
%If the Mott plateau is well developed, the shoulder structure thus appears in the curve of $S/N_{\rm tot}$.

Grand potential also provides us with useful information on the Mott transition of the present system.
As shown in Fig. \ref{fig_ent}(b), when $U/W=1.5$ ($2.0$), the gradient of $\Omega/N_{\rm tot}$ curves gradually changes around $E_t/W\sim0.75$ ($1.0$) indicating that there is the crossover between the Mott and metallic phases.
Here, the Mott (metallic) phase is defined as the phase in which the Mott insulating (metallic) region is dominant. %, i.e. the Mott plateau is (not) well developed.
We find that, in the Mott and metallic phases, the grand potential obeys the relations $\Omega\propto -E_t N_{\rm tot}$ and $\propto -E_t N_{\rm tot}/2$, respectively.
Note that the characteristic trap energy $E_t$ corresponds to the energy required to add one atom around the edge of the atomic cloud, namely the chemical potential in the limit of zero-tunneling.
% which is the energy requied to add one atom around the edge of the atomic cloud .
The difference between these relations, a factor $2$, results from the fact that double energy is required to add one atom in the Mott phase. 
% due to the suppression of the doubly occupied site in the Mott phase.

%%%===%%%  r1  結果の導入 モット絶縁体：温度依存性
%...................................
\begin{figure}[t]
\includegraphics[width=8cm]{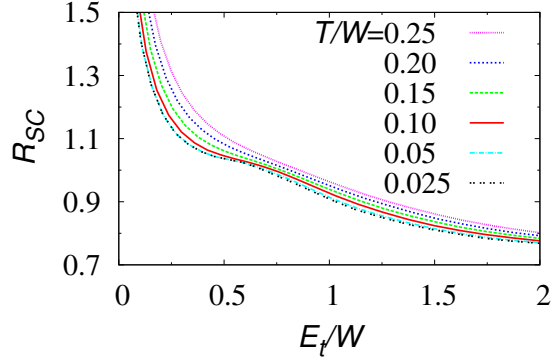}
\caption{(Color online) Rescaled cloud size $R_{SC}$ vs $E_t/W$ for different $T/W$ values.
 The interaction strength is fixed at $U/W=1.5$.}
\label{fig_rsc_T}
\end{figure}
\begin{figure}[t]
\includegraphics[width=8cm]{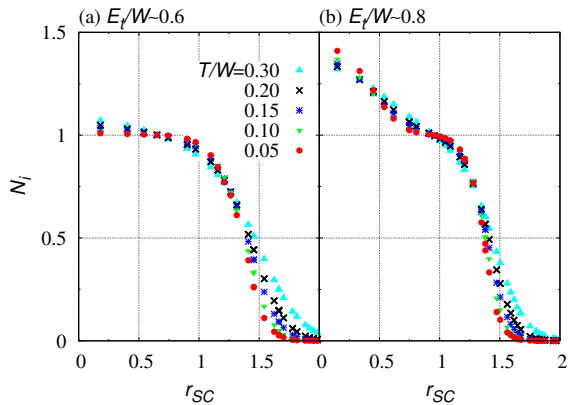}
\caption{(Color online) Density profiles of atomic clouds for two characteristic trap energies: (a) $E_t/W\sim0.6$ and (b) $E_t/W\sim0.8$. The temperature is varied from $T/W=0.05$ to $T/W=0.3$. The interaction strength is fixed at $U/W=1.5$.}
\label{fig_num_Tdep}
%,{\it i.e.} the number of atoms $N_i$ as a function of the rescaled distance $r_{SC}$,
\end{figure}
%...................................
We now shift our attention to the effects of thermal fluctuations on the quantities shown above.
In Fig. \ref{fig_rsc_T}, we show the temperature dependence of the $R_{SC}$ curves for $U/W=1.5$.
Figure \ref{fig_num_Tdep} shows the temperature dependence of the density profiles for $U/W=1.5$ at $E_t/W\sim0.6$ and $1.0$.
%%%===%%%  r2  発見の提示  モット絶縁体： 温度依存性　Rsc(Et)

From Fig. \ref{fig_rsc_T}, we find that, below $T/W=0.10$, $R_{SC}$ curves have a shoulder structure around $E_t/W\sim 0.5$, and change little as $T/W$ is decreased.
On the other hand, above $T/W=0.10$, thermal fluctuations destroy the shoulder structure; therefore, we find no signature of the Mott insulating state in the $R_{SC}$ curves.
However, as shown in Fig. \ref{fig_num_Tdep}, the Mott plateau survives up to $T/W\sim0.15$, suggesting that the destruction of the shoulder structure of $R_{SC}$ curves is mainly attributed to the thermal excitation of metallic atoms at the edge of the Mott insulating region.
%because the atoms in the Mott region are stable against thermal fluctuations due to the Mott gap.
%Below this characteristic temperature $T^*/W(\sim0.15)$, the density profile except for the edge of the Mott insulating region hardly changes with decreasing temperature.
%Consequently, the destruction of the shoulder structure is mainly attributed to thermal excitations of metallic atoms at the edge of the Mott insulating region. 
Consequently, for $E_t/W\sim0.6$ and $0.8$, we roughly estimate a specific temperature $T^*/W$ value of $\sim0.15$ around which the Mott plateau is destroyed by thermal fluctuations.

%%%===%%%  r1  結果の導入 モット絶縁体：温度依存性 S(Et)
%...................................
\begin{figure}[t]
\includegraphics[width=8cm]{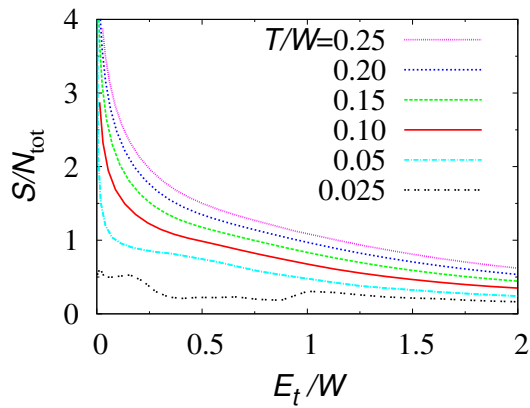}
\caption{(Color online) Entropy per atom $S/N_{\rm tot}$ vs $E_t/W$ for several $T/W$ values. The interaction strength is fixed at $U/W=1.0$.}
\label{fig_ent_T}
\end{figure}
%...................................
We also calculate the entropy per atom $S/N_{\rm tot}$ by varying the temperature as shown in Fig. \ref{fig_ent_T}.
%%%===%%%  r2  発見の提示   モット絶縁体：温度依存性　S(Et) 
%In the same way as above discussions, thermal fluctuations destroy the shoulder structure.
The $S/N_{\rm tot}$ curves do not saturate at low temperatures ($T/W\lesssim0.1$) in contrast to those of $R_{SC}$.
At the lowest temperature $T/W=0.025$ in Fig. \ref{fig_ent_T}, the entropy per atom is highly suppressed in the region $0.4\lesssim E_t/W \lesssim 0.8$.
This is due to the fact that a magnetic ordered phase appears at $T/W=0.025$, as discussed in detail in the next section.
We note that entropy is sensitive to magnetic transitions.

%%%===%%%  r3  関連した研究 モット絶縁体について
As shown in this section, SFA allows us to calculate various thermodynamic quantities at finite temperatures.
%In particular, it is difficult to obtain the entropy and the grand potential by using R-DMFT\cite{Snoek2008}.
Such thermodynamic quantities have already been investigated experimentally in ultracold atomic gases trapped in a conventional magnetic trap \cite{Ensher1996,Mewes1996,Jin1996}.
We hope that these thermodynamic quantities will be measured in optical lattice systems in the near future.
%These thermodynamic quantities have much useful information of the properties of atoms in an optical lattice.
%%%===%%%  r4  コメント モット絶縁体について
%Our results may pave the way for the study of the thermodynamic quantities, such as the entropy, in the optical lattice system.
%This point is discussed later in the following sections.

%%%%-------------------------------------------------------------%%%%
%%%%%　　　%%　　　%%%　　　%%　　　　%%　%%%　　　　%%%　%%%%　%%%%%
%%%%%　%%%%%%　%%%%%%　%%%%%%%%%%　%%%%%　%%%　%%%%　%%%　　%%　%%%%%
%%%%%%　　%%%　　　%%　%%%%%%%%%%　%%%%%　%%%　%%%%　%%%　%　%　%%%%%
%%%%%%%%%　%%　%%%%%%　%%%%%%%%%%　%%%%%　%%%　%%%%　%%%　%%　　%%%%%
%%%%%　　　%%　　　%%%　　　%%%%%　%%%%%　%%%　　　　%%%　%%%%　%%%%%
%%%%-------------------------------------------------------------%%%%
\section{Magnetic ordered phase}\label{sec_AF}
%%%===%%%  r1  結果の導入 磁性
%...................................
\begin{figure}[t]
\includegraphics[width=8cm]{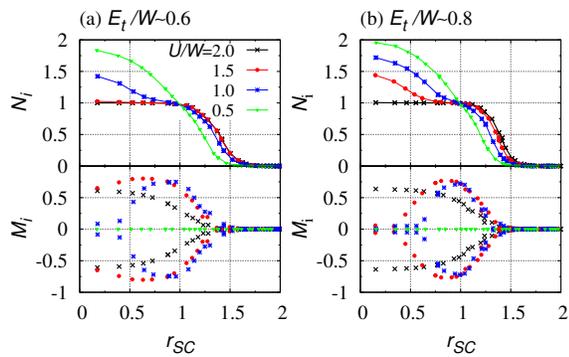}
\caption{(Color online) ({\it top}) the magnetization $M_i$ and ({\it bottom}) the number of atoms $N_i$ as functions of the rescaled distance $r_{SC}$ for two characteristic trap energies: (a)$E_t/W\sim0.6$ and (b)$E_t/W\sim0.8$.
The interaction strength is varied from $U/W=0.5$ to $U/W=2.0$ and the temperature is fixed at $T/W=0.025$. }
\label{fig_mag}
\end{figure}
%...................................
\begin{figure}[t]
\includegraphics[width=8cm]{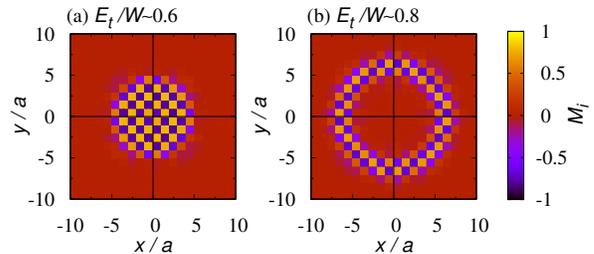}
\caption{(Color online) Real-space distribution of the magnetization at $U/W=1.5$ and $T/W=0.025$ for (a)$E_t/W\sim0.6$ and (b)$E_t/W\sim0.8$. 
%Here, $x$ ($y$) is the vertical (horizontal) distance from the center of the trap. %, where its unit is a lattice distance.
}
\label{fig_mag_ii}
\end{figure}
%...................................
In the previous section, we showed that entropy shows a characteristic of the AF magnetic transition at a low temperature $T/W=0.025$.
Here, to clarify the properties of the magnetic ordered phase, we investigate the present model at a fixed temperature $T/W=0.025$.
%First, we show that the magnetization of atoms in real-space shows a good agreement with the previous result at zero temperature in Ref. \cite{Snoek2008}.
%Next, by systematic calculation with varying the temperatures, we demonstrate how the magnetic ordered phases develop in our model.
%In addition, we find that the behavior of the entropy as a function of temperatures indicates the second or higher order magnetic transition.
%We end up with the estimation of the transition temperatures.

We first look at the magnetization $M_i=\langle n_{i\uparrow}\rangle-\langle n_{i\downarrow}\rangle$.
In Fig. \ref{fig_mag}(a) and (b), we show $M_i$ and $N_i$ as functions of $r_{SC}=r_i/r_t$ for different $U/W$ values at $E_t/W\sim0.6$ and $0.8$, respectively.
%%%===%%%  r2  発見の提示   磁性 M(r) with Et
Except for a weakly interacting case of $U/W=0.5$, we find finite values of $M_i$.
The sign of $M_i$ changes alternately from site to site, suggesting the AF order.
To clarify this point, we also show the real-space distribution of the magnetization $M_i$ for $U/W=1.5$ in Fig. \ref{fig_mag_ii}.
We note that the rotational symmetry is broken in the AF ordered region.
% in our system in which the center of trapping potential is set at 
The magnetization is well-developed in the Mott plateau region.
In addition, even in the metallic region in the vicinity of the Mott plateau, the magnetization $M_i$ has a small but finite value \cite{Snoek2008}.

From Figs. \ref{fig_num_Tdep} and \ref{fig_mag}, we see that the density profiles of atoms remain unchanged at very low temperatures even though the magnetization $M_i$ becomes finite.
As a consequence, the appearance of a magnetic ordered region hardly affects the rescaled cloud size $R_{SC}$ (see also Fig. \ref{fig_rsc_T}).

%%%===%%%  r3  関連した研究 磁性
The magnetic ordered phase has already been investigated at zero temperature using R-DMFT by M. Snoek {\it et al.} in Ref. \cite{Snoek2008}.
Our results at very low temperature agree well with their results at zero temperature.
%%%===%%%  r4  コメント 磁性
%However, to evaluate the critical temperatures of magnetic transition in the optical lattice systems, it is necessary to take account of the effects of thermal fluctuations.
We next focus on the temperature dependence of magnetization, and evaluate the critical temperatures of the AF magnetic transition in the present system.

%%%===%%%  r1  結果の導入 磁性：温度依存性 M(T)
%...................................
\begin{figure}[t]
\includegraphics[width=8cm]{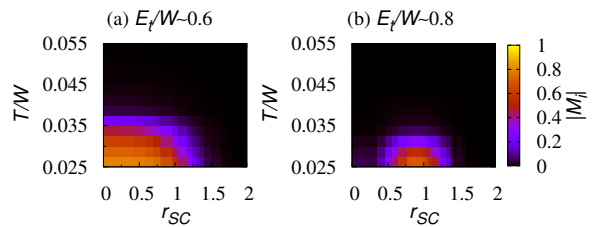}
\caption{(Color online) Absolute value of the magnetization $|M_i|$ as a function of $r_{SC}$ and $T/W$ for fixed values of (a)$E_t/W\sim0.6$  and (b)$E_t/W\sim0.8$.}
\label{fig_mag_Tdep}
\end{figure}
%...................................
We systematically calculate the magnetization by varying the temperature.
Figure \ref{fig_mag_Tdep}(a) and (b) show the absolute values of $M_i$ for $U/W=1.5$ at $E_t/W\sim0.6$ and $0.8$, respectively.
%%%===%%%  r2  発見の提示 磁性：温度依存性
The magnetic ordered region gradually spreads as $T/W$ decreases.
% and finally vanishes at $T/W\sim0.033$ (0.032) for $E_t/W\sim 0.6$ and $0.8$.
Thus we can understand the quantum phase transitions in the present system as follows.
The Mott plateau first develops at higher temperatures $T^*/W(\sim0.15)$ (see Fig. \ref{fig_num_Tdep}), and then the magnetic ordered region gradually grows from the inside of the Mott plateau as the temperature decreases below the transition temperature $T_c/W(\sim0.04)$.
%It results from the fact that the atoms in the Mott plateau region is stable aginst temperature due to the Mott gap.
Note that a well-developed Mott plateau region ($E_t/W\sim0.6$) has higher transition temperatures than those of the shell-like region ($E_t/W\sim0.8$).
%We find that the magnetic ordered region continuously develops with decreasing $T/W$.

%%%===%%%  r1  結果の導入 磁性：温度依存性 S(T)
%...................................
\begin{figure}[t]
\includegraphics[width=8cm]{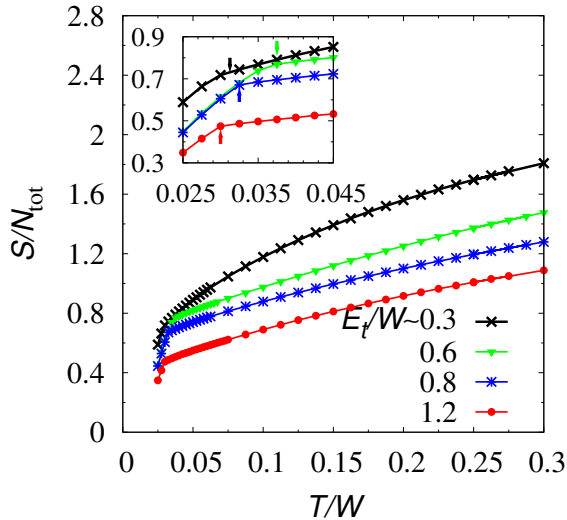}
\caption{(Color online) Entropy per atom $S/N_{\rm tot}$ as a function of $T/W$ for several values of $E_t/W$ at $U/W=1.5$.
Inset is an enlarged view and arrows indicate the magnetic transition temperatures.
}\label{fig_ent_vs_T}
\end{figure}
%...................................
For a more precise discussion, we calculate the temperature dependence of the entropy per atom.
In Fig. \ref{fig_ent_vs_T}, we show $S/N_{\rm tot}$ as a function of $T/W$ at $U/W=1.5$.
%%%===%%%  r2  発見の提示 磁性：温度依存性 S(T)
For $E_t/W\sim1.2, 0.8, 0.6$ and $0.3$, the $S/N_{\rm tot}$ gradient suddenly becomes steeper at the transition temperatures $T_c/W=0.029, 0.038, 0.048$ and $0.031$, respectively.
%It indicates that the Mott plateau develops in this region.
% where the magnetization becomes finite value as shown in Fig. \ref{fig_mag_Tdep}.
In the region from just above the transition temperatures $T_c/W$ to the specific temperatures $T^*/W$, we find a linear temperature dependence of $S/N_{\rm tot}\propto T$.
Significantly, except for the small $E_t/W$ value of $\sim0.3$, the curves of the entropy per atom show a cusp-like anomaly at $T_c/W$, suggesting second order magnetic transitions.
%It suggests that the magnetic ordered transition in the system with the confinement may be of the second order.
However, we cannot definitely determine the order of the transition in our present calculation.
We reach the conclusion that the transition is of the second or higher order.

Let us discuss this temperature dependence of the entropy per atom in more detail.
In a uniform system ($V_t=0$), the linear temperature dependence of entropy $S\propto T$ is a characteristic of a metallic state, while a constant entropy $S=\ln2$ is a characteristic of the Mott insulating state at which a localized free spin is induced at each site.
On the other hand, in a trapped system ($V_t\not=0$), both states coexist below $T^*$, namely the characteristic temperature of a well-developed Mott plateau;
therefore, in the region $T_c<T<T^*$, the entropy per atom obeys the relation $S/N_{\rm tot}=bT+c\ln2$, where $b$ and $c$ are, roughly speaking, the coefficient depending on the density profiles and the effective mass of atoms.
%where $b$ is a coefficient depending on the fraction of atoms in the Mott insulating region and $a$ is that depending on the effective mass of atoms in the matallic region.
As the temperature is decreased below $T^*$, free spins in the Mott insulating region interact with each other, and then order antiferromagnetically at $T=T_c$ where the redundant entropy $c\ln2$ is suddenly released.
Above $T^*$, a complicated temperature dependence can be seen that results from thermal fluctuations of the atoms in the Mott insulating region.
Note that the entropy, which is enlarged by strong correlations, and its release play key roles for the magnetic transitions.
Therefore, we first observe the developed Mott plateau around $T^*$, and then find that the magnetic ordered region appears inside this plateau at $T_c$.
%At higher temperatures, as decreasing temperatures, localized free spins are induced in the Mot insulating region by the strong local correlations.

%%%===%%%  r1  結果の導入 相図

\begin{figure}[th]
\includegraphics[width=8cm]{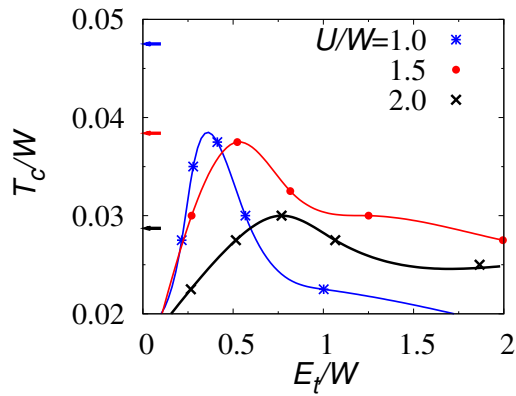}
\caption{(Color online) AF transition temperatures $T_c/W$ vs $E_t/W$ for different $U/W$ values. 
The lines are guides for the eyes. 
The arrows indicate AF transition temperatures for the uniform Hubbard model at half-filling for $U/W=1.0, 1.5$ and 2.0 from top to bottom.}
\label{fig_phase}
\end{figure}

Next, we systematically estimate the AF transition temperatures $T_c/W$ for several choices of $U/W=1.0, 1.5$ and $2.0$.
The results are shown in Fig. \ref{fig_phase}.
%%%===%%%  r2  発見の提示   相図　Tc(Et)
In each curve, the transition temperatures as a function of $E_t/W$ have a maximum value around $E_t/W\sim 0.5$ where we see the well-developed Mott plateau.
When $E_t/W$ is small, there are insufficient atoms to form a magnetic ordered phase, leading to a very small $T_c/W$.
The long tail of $T_c/W$ curves in the large $E_t/W$ region results from the fact that the shell-like structure of the Mott plateau survives in such regions.
For comparison, we calculate $T_c^{\rm uni}$: the AF transition temperatures for the uniform Hubbard model at half-filling ($V_t=0$ and $N_{\rm tot}=L$). 
We show $T_c^{\rm uni}/W$ for $U/W=1.0, 1.5$ and $2.0$ by the arrows in Fig. \ref{fig_phase}.
With a strongly interacting limit, $T_c^{\rm uni}$ is inversely proportional to $U$, while for a weakly interacting limit, $T_c^{\rm uni}$ decreases exponentially with decreasing $U$ \cite{Jarrell1992}. % ($T_c^{\rm uni}\propto 1/U$)
Indeed, $T_c^{\rm uni}/W$ has a maximum value around $U/W\sim1.0$.
%, and gradually decrease with increasing $U/W$.
%We show the transition temperatures of uniform system as arrows in Fig. \ref{fig_phase}.
On the other hand, in the system with a trapping potential, the transition temperatures for $U/W=1.0$ are comparable to those for $U/W=1.5$.
This is because the Mott plateau does not develop at $U/W=1.0$, because the effects of the confinement prevail against those of the correlations.
For $U/W=1.5$, we find higher transition temperatures than those for $U/W=1.0$ and $2.0$ over a wide $E_t/W$ range.
%%%===%%%  r3  関連した研究 相図
%%%===%%%  r4  コメント 相図
%By systematic analyses taking account of the effects of thermal fluctuations, correlations, and a confinement potential, 
The results in Fig. \ref{fig_phase} suggest that the system for $U/W\sim1.5$ is more suitable for an observation of the magnetic ordered phase.
%The entropy per atom cleary shows the evidence of the magnetic ordered phase.***

Before concluding this section, we compare our present results to other numerical works which have discussed the AF transition temperatures $T_c$ in optical lattice systems without consideration of the effects of the trapping potential \cite{Hofstetter2002,Werner2005}.
F. Werner {\it et. al} have estimated $T_c/W\sim0.04$ at $U/W\sim 1$ \cite{Werner2005}, and W. Hofstetter {\it et. al} have done $T_c/W\sim0.04$ at $U/W=0.5$ \cite{Hofstetter2002}. 
Our estimated value of $T_c/W\sim0.04$ around $E_t/W\sim 0.5$ at $U/W=1$ shows a reasonable agreement with these previous results.
We note that, however, lower $T_c$ values are realized except for around $E_t/W\sim 0.5$ in Fig. \ref{fig_phase}.
Furthermore, for a weakly interacting case $U/W=0.5$, we cannot find the AF transition down to $T/W=0.02$ due to the effects of the trapping potential. 

%...................................
%...................................

%%%%-------------------------------------------------------------%%%%
%%%%%　　　%%　　　%%%　　　%%　　　　%%　%%%　　　　%%%　%%%%　%%%%%
%%%%%　%%%%%%　%%%%%%　%%%%%%%%%%　%%%%%　%%%　%%%%　%%%　　%%　%%%%%
%%%%%%　　%%%　　　%%　%%%%%%%%%%　%%%%%　%%%　%%%%　%%%　%　%　%%%%%
%%%%%%%%%　%%　%%%%%%　%%%%%%%%%%　%%%%%　%%%　%%%%　%%%　%%　　%%%%%
%%%%%　　　%%　　　%%%　　　%%%%%　%%%%%　%%%　　　　%%%　%%%%　%%%%%
%%%%-------------------------------------------------------------%%%%
\section{Summary}\label{sec_summary}
%%%===%%%  d1  問題と発見の再確認 
We have investigated the two-component fermionic atoms on a two-dimensional (2D) optical lattice with a harmonic trapping potential at finite temperatures.
%%%===%%%  d2  重要性の強調   
For a comprehensive understanding of both the Mott transition and the magnetic transition in this system, we have extended the self-energy functional approach (SFA) to deal with inhomogeneous systems.
By introducing additional self-consistent loops, a complicated variational problem for an inhomogeneous system following the framework of the SFA is decomposed into several easily solvable SFA problems.
%This mehod allows us to systematically investigate various quantities which could be experimentally observed.
%we have to appropriately take account of all effects of strong correlations, a tunneling term, thermal fluctuations, and a trapping potential.
%%%===%%%  d3  過去の研究との差異 
%%%===%%%  d4  主要な主張   
We have applied this method to the 2D Hubbard model with a trapping potential.
A proper rescaling of the system parameters allows us to qualitatively compare the two-dimensional system with the three-dimensional (3D) one.
The calculated results of the rescaled cloud size and density profiles show good qualitative agreement with previous results in Ref. \cite{Schneider2008}.
Furthermore, we have systematically calculated thermodynamic quantities such as entropy and grand potential.
We clarified that entropy shows evidence of both the Mott transition and the antiferromagnetic (AF) transition.
In addition, we have demonstrated how confinement affects the AF transition temperature.
We have estimated the AF transition temperature, and proposed a suitable parameter region for observing the AF ordered phase experimentally.

A direct comparison with experimental results in the 2D optical lattice systems will be useful. 
The detail information of experimental setups is required for such studies. 
Additionally, it is very important to investigate $d$-wave superfluid phase \cite{Hofstetter2002}. 
However, dealing with the $d$-wave superfluid correlations is not straightforward in our present choice of the reference system. 
These issues are beyond our current scope and left as important future works.
%%%===%%%  d5  研究の適用範囲と限界 

\begin{acknowledgments}
We thank Y. Takahashi, A. Koga, S. Suga and Y. Tokura for valuable discussions.
\end{acknowledgments}

%%%%%%%%%%%%%%%%%%
%%%%%%%%%%%%%%%%%%
%%%%%%%%%%%%%%%%%%
\appendix
\section{Role of the atomic bath}
%...................................
\begin{figure}[th]
\includegraphics[width=8cm]{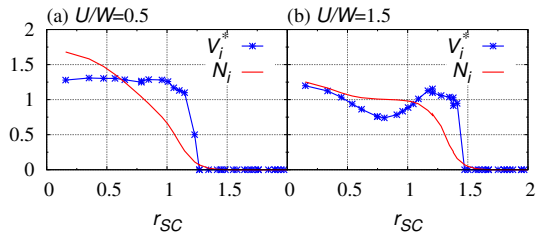}
\caption{
(Color online) The optimized hybridization $V_i^*$ as a function of $r_{SC}$ for two choices of the interaction strength: (a) $U/W=0.5$ and (b) $U/W=1.5$. Other parameters are fixed at $T/W=0.05$ and $E_t/W\sim0.8$.
For convenience, we also plot the number of atoms $N_i$.
}
\label{fig_hyb}
\end{figure}
%...................................
In this appendix, we comment on the role and importance of the atomic bath with the hybridization $V_{i\sigma}$ in the reference Hamiltonian of Eq. (\ref{eq_ref_model}). 

As mentioned in Sec. \ref{sec_method}, we adopt the $L$-sets of two-site Anderson impurity models as the reference system. Here, we explain this procedure in more detail.
We first separate a lattice into localized single sites.
%, namely we set $J=0$ in Eq. (\ref{eq_model}).
Then, for each separated site, we additionally introduce the site consisting of non-interacting atoms with the hybridization, and set the energy level of such sites just at Fermi level of the original system.
Accordingly, these non-interacting sites are partially occupied and play a role of the atomic reservoir.
Furthermore, the hybridization in the reference Hamiltonian corresponds to the hopping in the original Hamiltonian.

It is known that, when the interaction strength is increasing, an effective energy scale of the hopping becomes much smaller.
In other words, the renormalization effects are induced by the effects of correlations.
We will demonstrate that the hybridization $V_{i\sigma}$ in the reference Hamiltonian clearly reflects such many-body dynamics via the variational condition $\partial \Omega/\partial V_{i\sigma}=0$. %(\ref{eq_v_condition}). 
In Fig. \ref{fig_hyb}, we plot the optimized values of the hybridization $V^*_{i\sigma}$ which satisfy all of the required conditions mentioned in Sec. \ref{sec_method}. 
%$\partial \Omega/\partial V_{i\sigma}=0$. % with $\left[{\bf G}_{\sigma}\right]_{ii}=G^{\rm loc}_{i\sigma}$.
Figure \ref{fig_hyb}(a) and (b) show $V_i^*\equiv V^*_{i\uparrow}=V^*_{i\downarrow}$ as a function of $r_{SC}$ for $U/W=0.5$ and $1.5$, respectively. 
We set other parameters as $T/W=0.05$ and $E_t/W\sim0.8$.
In Fig. \ref{fig_hyb}(a), we see large and constant $V_i$ values in an occupied region ($r_{SC} \lesssim 1.3$).
Note that, we reasonably choose $V_{i}=0$ in an unoccupied region ($r_{SC}\gtrsim 1.3$) since the variational conditions $\partial \Omega/\partial V_{i\sigma}=0$ is satisfied by any values of $V_{i\sigma}$. 
On the other hand, in Fig. \ref{fig_hyb}(b), we see that values of $V_i$ decrease in the Mott plateau region ($0.5<r_{SC}<1$), indicating the renormalization effects.

%===============================================

%\bibliography{../../../../sfa,../../../../published,../../../../papers,../../../../reviews}

\end{document}